\documentclass[final,5p,times,twocolumn]{elsarticle}

\usepackage{amssymb}

\journal{Physica E}

\begin{document}

\begin{frontmatter}



\title{Gap closing and universal phase diagrams in topological insulators}


\author[label2,label3]{Shuichi Murakami}

\address[label2]{Department of Physics, Tokyo Institute of Technology, 
2-12-1 Ookayama, Meguro-ku, Tokyo 152-8551, Japan}
\address[label3]{PRESTO, 
Japan Science and Technology Agency (JST),
4-1-8 Honcho, Kawaguchi,
Saitama 332-0012 Japan}

\begin{abstract}
We study a general problem
how the gap in a nonmagnetic band insulator closes 
by tuning a parameter. We review our 
recent results on the classification of 
all the possible gap closing in two and three dimensions.
We show that they accompany the change of Z$_2$ topological 
numbers, and that the gap closings correspond to
phase transitions between the quantum spin Hall and the 
insulator phases. 
Interestingly, in inversion-asymmetric three-dimensional systems there 
appears a gapless phase between the quantum spin Hall and 
insulator phases.
This 
gapless phase is due to a topological nature of gap-closing points in three dimensions, but not in two dimensions.

\end{abstract}

\begin{keyword}
Topological insulator \sep topological number \sep quantum spin Hall effect
\sep spin current 

\end{keyword}

\end{frontmatter}


\section{Introduction}
The physics of pure spin current 
has been attracting much attention recently in various fields of condensed matter.
Pure spin current is a flow of electron spins which is not accompanied by the charge current.
As is different from the spin itself, 
spin current is even under the time-reversal operation, 
and thus can be nonzero in nonmagnetic systems. 
Thereby the spin current has opened up new fields in nonmagnetic condensed materials. 
One example is 
the intrinsic spin Hall effect (SHE) \cite{Murakami03a,Sinova04} driven by the spin-orbit interaction.
The SHE has offered us a new way to inject and detect spin currents in metals and semiconductors. 

The physics of topological insulators, also called the quantum spin Hall (QSH) systems \cite{Kane05a,Kane05b,Bernevig05a}, 
is another example
of new physics due to the pure spin current. Two-dimensional (2D)  QSH systems are insulators in the bulk, 
while they have gapless edge states. Similarly, three-dimensional (3D) QSH systems (topological insulators) are insulators with gapless surface states. One of the important and novel aspects of these
systems is that these gapless edge/surface states are robust against perturbation which respects 
time-reversal symmetry. Namely, they remain gapless even when nonmagnetic impurities or disorder exist 
\cite{Wu05,Xu05}, which causes novel transport properties \cite{Konig07,Takahashi10}.
The robustness of these gapless states can be interpreted as protected by topology. The 
QSH systems are characterized by nontrivial Z$_2$ topological number.
The $Z_2$ topological number takes only 
two different values for $\nu$: $\nu=0$ (also called $\nu=$even) or 
$\nu=1$ (also called $\nu=$odd).
When $\nu=1$, the system is in the QSH phase, while
when $\nu=0$ the system is in the insulating phase.  This topological number remains unchanged as long as the bulk gap is open and the time-reversal symmetry is preserved. Transitions between ordinary insulator ($\nu=0$) and 
the QSH system ($\nu=1$) occur only when the gap is closed.
The topological classification has been discussed in various theoretical papers, but they might not be easily 
accessible for non-experts in topology. 

Instead of directly dealing with the Z$_2$ topological numbers, 
we take a different route and study the physics of gap-closing. This has been 
studied in a series of papers of the author \cite{Murakami07a,Murakami07b,Murakami08}.
In the present paper, we summarize these results in these papers.
In generic nonmagnetic band insulators, 
we examine whether the gap closes when we 
vary a parameter in the system. 
As a result we can classify all the 
possible types of gap-closing. In fact, in every case
of the gap closing, we show in this paper 
that the Z$_2$ topological number changes.  Namely,
the gap-closing physics is equivalent to 
the phase transition between the QSH system and the ordinary insulator. As a byproduct, we
obtain effective theories focusing on the vicinity of the gap-closing points at the transition.
This gap-closing physics involves local features in ${\bf k}$ space, and is equivalent to global topological
structure in ${\bf k}$ space \cite{Kane05b,Fu06a,Fu06c,Fu07a,Moore07}.
We note that this gap-closing physics tells us only the {\it change} of the topological number, and we cannot tell which side of the gap-closing is 
the topological insulator and which side is thee trivial insulator. 
This is the characteristic of the theories involving local features in 
${\bf k}$ space. On the other hand, the
gap-closing physics gives us a number of new results, such as effective theories 
expanded in terms of $k$, or the universal phase diagrams, as we discuss 
in this paper.

As a result of our analysis, we get a full understanding of the gap-closing physics,
which is different between 2D \cite{Kane05a,Kane05b,Bernevig05a} and 3D 
\cite{Fu07a,Moore07}, and 
between inversion-symmetric and inversion-asymmetric systems. 
The most interesting is the 3D inversion-asymmetric systems; 
we find that there necessarily exists a finite region of 
gapless phase between the 
QSH phase and the ordinary insulator phase. This gapless phase is
of topological origin, as we describe in this paper. 
Henceforth we consider only clean systems without any impurities or
disorder. The time-reversal symmetry is
assumed throughout the paper.
In the present paper we only deal with time-reversal-invariant systems with spin-orbit coupling, 
without assuming additional symmetries such as spin-rotational symmetry. 
Therefore, the systems without spin-orbit coupling, such as the organic
system with bulk Dirac-like bands \cite{Kobayashi07}, 
belong to a different class of systems (orthogonal class)
and are beyond the scope of the present paper.

\section{Gap closing at the phase transition}
Putting topological insulators aside for a while, 
we ask ourselves 
a general question when and how the gap of a nonmagnetic band insulator 
(with spin-orbit coupling) closes by tuning a {\it single} parameter $m$ in the system. 
We assume that the system is time-reversal symmetric.
Depending on the system considered, this parameter $m$ can have any physical meaning, such as pressure, chemical content, or interatomic distance.
Because our goal is to pursue the phase transition, 
we will focus on  gap-closing achieved by tuning
 a {\it single} parameter $m$ in generic systems. We call such gap closing
as ``generic gap closing". 
The word ``generic" means that any terms allowed by symmetry are set to be nonzero. There are several reasons why we focus on generic gap closing. First, in real materials there are many factors which determine the band structure. Therefore, any kind of terms in the Hamiltonian are set to be nonzero, as long as the terms are allowed by symmetry. In other words, we assume that there are no accidental degeneracies 
unless it is required by symmetry.
Second, only the {\it generic} gap closings are necessarily related with phase transitions. Nongeneric ones do not correspond to phase transition, because such kind of nongeneric gap-closing may disappear by small perturbation. 
(If such nongeneric gap closing does not disappear by perturbation, it is nothing but a generic gap closing studied in this paper. Thus it is enough to study only the generic gap closing.)
By these reasons, non-generic gap-closing which requires fine tuning of some other parameters is excluded in our analysis.

With these reasonable and general assumptions, we consider a problem whether the 
gap will close by tuning only a single parameter $m$ in the system. 
In respective cases considered in the following, we find necessary conditions for the gap to close, and see whether it can be satisfied by changing a parameter $m$.
We consider 
a gapped spin-$1/2$ time-reversal-symmetric system with spin-orbit interaction.
A Hamiltonian matrix for Bloch wavefunctions can be written in 
a block form,
\begin{equation}
H({\bf k})=\left(\begin{array}{cc}
h_{\uparrow\uparrow}({\bf k})& h_{\uparrow\downarrow}({\bf k})\\
h_{\downarrow\uparrow}({\bf k})& h_{\downarrow\downarrow}({\bf k})
\end{array}\right).\label{eq:Hamiltonian}
\end{equation}
The dimensions of the Hamiltonian $H({\bf k})$
can be arbitrary for the subsequent analysis, but for simplicity
 we set the dimension to be minimal, as long as the gap-closing 
physics is appropriately captured; we retain only the 
states which are involved in the gap closing. 
In the following, we find that the 
dimension turns
out to be four (see \S \ref{sec:I-sym}) and two (see \S \ref{sec:I-asym}) for the inversion-symmetric and inversion-asymmetric systems, 
respectively.
The spectrum 
is assumed to have a gap, within which 
the Fermi energy lies. 
Because the time-reversal operator $\Theta$ is given by $\Theta=i\sigma_y K$ 
with $K$ being complex conjugation, the
time-reversal-symmetry results in 
\begin{equation}
H({\bf k})=\sigma_y  H^{T}(-{\bf k})\sigma_y,
\label{time-reversal}\end{equation}
which is cast into the equations  
$h_{\uparrow\uparrow}({\bf k})
=h_{\downarrow\downarrow}^{T}(-{\bf k})$ and
$h_{\uparrow\downarrow}({\bf k})
=-h_{\uparrow\downarrow}^{T}(-{\bf k})$.

Before proceeding to the specific cases, we introduce the 
momenta with 
${\bf k}\equiv -{\bf k}$ (mod ${\bf G}$) 
where 
${\bf G}$ is a reciprocal lattice vector.
Such momenta are called 
the time-reversal invariant momenta (TRIM) ${\bf \Gamma}_i$
\cite{Kane05b,Fu06a,Fu06c,Fu07a,Moore07},
and have the values ${\bf \Gamma}_{i=(n_1n_2n_3)}=(
n_1{\bf b}_{1}
+n_2{\bf b}_{2}
+n_3{\bf b}_{3})/2$  in 3D, and 
${\bf \Gamma}_{i=(n_1n_2)}=(
n_1{\bf b}_{1}
+n_2{\bf b}_{2})/2$ in 2D, where
$n_j=0,1$ and ${\bf b}_{j}$ are primitive reciprocal lattice vectors.  
These momenta are invariant under time-reversal, and are important in the subsequent discussion. 

Before proceeding to the cases with inversion-symmetric and inversion-asymmetric
systems, we explain the definition of inversion symmetry. 
The space inversion means that the sign of every space coordinate is changed, and 
the crucial point here is that the inversion leaves spins unchanged.
In three dimensions it means $x\rightarrow -x$, $y\rightarrow -y$, and
 $z\rightarrow -z$. 
In two dimensions the space coordinates change as 
$x\rightarrow -x$, $y\rightarrow -y$, while it is different from the 
rotation within $xy$ plane, because the space rotation changes $s^x\rightarrow 
-s^x$, 
$s^y\rightarrow 
-s^y$, and $s^z\rightarrow 
s^z$, while inversion does not change spin.
To see this more clearly, we take the Rashba spin-orbit term $\lambda({\bf s}\times
{\bf k})_z$ as an example, where ${\bf s}$ is the spin. 
This term changes sign under inversion since ${\bf k}$ 
flips its sign while ${\bf s}$ does not. 
This sign change under inversion can be regarded as a reversal of the $+z$
direction, which
represents structural inversion  asymmetry causing the Rashba coupling.

\subsection{Inversion symmetric systems}
\label{sec:I-sym}
In the inversion-symmetric systems, the energies 
are doubly degenerate
at each ${\bf k}$. The gap closing involves a 
doubly-degenerate valence band and a doubly-degenerate conduction band. 
Hence, four states are involved in gap closing, and we consider 4$\times$4 Hamiltonian 
matrix 
$H({\bf k})$ ($h_{\alpha \beta}({\bf k}$) in 
Eq.~(\ref{eq:Hamiltonian}) is 2$\times$2). 
The inversion-symmetry is expressed as
\begin{equation}
H(-{\bf k})=PH({\bf k})P^{-1}, \ u(-{\bf k})=Pu({\bf k}),
\label{eq:H-inversion}
\end{equation}
where $P$ is a unitary matrix independent of ${\bf k}$, 
and $u({\bf k})$ is the periodic part 
of the Bloch wavefunction: $\varphi_{{\bf k}}({\bf r})
=u({\bf k})e^{i{\bf k}\cdot{\bf r}}$.
As the inversion does not change spin, this 
unitary matrix $P$ is block-diagonal in spin space:
\begin{eqnarray}
P=\left(\begin{array}{cc}
P_{\uparrow}&\\&P_{\downarrow}
\end{array}
\right).\label{eq:P}
\end{eqnarray}
By using (\ref{eq:H-inversion}) and (\ref{eq:P}), after judicious unitary transformation,
all cases are shown to reduce to a case
$P_{\uparrow}=P_{\downarrow}={\rm diag}(\eta_{a},\ \eta_{b})$ with 
$\eta_{a}=\pm 1$, $\eta_{b}=\pm 1$.
 $\eta_a$ and $\eta_b$ represent the
parity eigenvalues of the atomic orbitals involved.

Occurrence of gap closing turns out to be different for the cases
(i) $\eta_a=\eta_b$ and 
(ii) $\eta_a=-\eta_b$.
The case (i) $\eta_{a}=\eta_{b}=\pm 1$ means that 
the atomic orbitals $a$, $b$ 
have the same parity, such as two $s$-like orbitals or two $p$-like 
orbitals. The inversion-symmetry (\ref{eq:H-inversion}) then imposes that  
$h_{\uparrow\uparrow}=h_{\downarrow\downarrow}^{T}$ is an even function 
of ${\bf k}$, and $h_{\uparrow\downarrow}=h_{\downarrow\uparrow}^{\dagger}$ is
an antisymmetric matrix, even function of ${\bf k}$.
The generic Hamiltonian 
becomes
\begin{eqnarray*}
&&{H}({\bf k})=E_{0}({\bf k})+\sum_{i=1}^{5}a_{i}({\bf k})\Gamma_{i}
 \\
&&=E_0+\left(\begin{array}{cccc}
a_3&a_1-ia_2 &0 &-a_4-ia_5 \\
a_1+ia_2& -a_3& a_4+ia_5 &0 \\
0& a_4-ia_5 &a_3 &a_1+ia_2  \\
-a_4+ia_5 &0 &a_1-ia_2& -a_3
\end{array}\right)
\end{eqnarray*}
where $a_{i}$'s and $E_{0}$ are real even functions of ${\bf k}$.
$\Gamma_{i}$ are $4\times 4$ matrices forming Clifford algebra, given by
$\Gamma_{1}=1 \otimes
\tau_{x}$, $\Gamma_{2}=\sigma_{z}\otimes\tau_{y}$, 
$\Gamma_{3}=1 \otimes\tau_{z}$, 
$\Gamma_{4}=\sigma_{y}\otimes\tau_{y}$, and  
$\Gamma_{5}=\sigma_{x}\otimes\tau_{y}$, where 
$\sigma_{i}$ and $\tau_{i}$ are Pauli matrices acting
on spin and orbital spaces, respectively.
The eigenenergies are given by 
$E_{0}\pm\sqrt{\sum_{i=1}^{5}a_{i}^{2}}$. 
The gap closes when $a_{i}({\bf k})=0$ for $i=1,\cdots,5$.
Namely, these five independent equations should be satisfied 
for the gap to close. This number is called codimension.
We see that in general there are no solutions of $k_x$, $k_y$ and $m$ satisfying these
five conditions. It follows 
from our assumption of ``generic" systems, i.e. we exclude a 
possibility that these five conditions are satisfied by accident by 
only three available parameters $k_x$, $k_y$ and $m$.
Hence, the gap never closes for (i) $\eta_a=\eta_b$. This 
is caused by the level repulsion between the states with the same parity.
As we can see from this example, when the codimension exceeds the number of 
available parameters, we cannot expect the gap closing to occur in generic systems.
Thus the codimension represents how hard it is to achieve gap closing, by overcoming the 
level repulsion between the conduction and the valence bands.

The result is different for the 
other case (ii) $\eta_{a}=-\eta_{b}=\pm 1$, i.e. $P=\eta_{a}
\tau_{z}
=\pm\tau_{z}$,
where the two atomic orbitals have different parity.
After a simple calculation, one can write the Hamiltonian as
\begin{equation}
{H}({\bf k})=a_{0}({\bf k})+a_{5}({\bf k})\Gamma'_{5}+\sum_{i=1}^{4}
b^{(i)}({\bf k})
\Gamma'_{i}
\end{equation}
where $a_0({\bf k})$ and $a_{5}({\bf k})$ are even functions of ${\bf k}$, and 
$b^{(i)}({\bf k})$ are odd functions of ${\bf k}$. Here
$\Gamma'_{i}$  are $4\times 4$ matrices forming Clifford algebra, given by  
$\Gamma'_{1}=\sigma_{z}\otimes\tau_{x}$,
$\Gamma'_{2}=1 \otimes\tau_{y}$,
$\Gamma'_{3}=\sigma_{x}\otimes\tau_{x}$,
$\Gamma'_{4}=\sigma_{y}\otimes\tau_{x}$,
and $\Gamma'_{5}=1 \otimes\tau_{z}$.
In this case the gap closes when five equations $a_{5}({\bf k})=0$,
$b^{(i)}({\bf k})=0$ are satisfied. 
At generic ${\bf k}$,
these five equations cannot be satisfied 
simultaneously even when a single parameter $m$ is changed, again by 
our assumption of generic systems.
On the other hand, at the TRIMs ${\bf k}={\bf \Gamma}_{i}$,  the situation is different. At these points ${\bf k}=-{\bf k}\ ({\rm mod} {\bf G})$ holds, and 
the odd functions $b^{(i)}({\bf k})$ vanish identically, 
and one has only to tune
$a_{5}({\bf k})$ to be zero. Thus, the gap closes at ${\bf k}={\bf \Gamma}_i$ 
by tuning a single parameter $m$.
By putting $\Delta{\bf k}\equiv {\bf k}-{\bf \Gamma}_i$, 
the Hamiltonian is expanded to linear order in $\Delta{\bf k}$ as
\begin{equation}
{H}({\bf k})\sim E_{0}+m\Gamma'_{5}+\sum_{i=1}^{4}
\left({\bf \beta}^{(i)}\cdot\Delta {\bf k}\right)
\Gamma'_{i},
\label{eq:Hb}
\end{equation}
where $E_0$ and $m$ are constants, and 
${\bf \beta}^{(i)}$ $(i=1,\cdots,4)$ are two-dimensional real constant vectors.
In this model, $m$ is the parameter which controls the gap closing, 
and the gap closes at the TRIM ${\bf \Gamma}_{i}$ when $m=0$.
After judicious 
unitary transformations,
the Hamiltonian finally becomes block-diagonal;
\begin{equation}
{H}({\bf k})=E_{0}+\left(
\begin{array}{cccc}
m&z_{-}&&\\
z_{+}&-m&&\\
&&m&-z_{+}\\
&&-z_{-}&-m
\end{array}\right).\label{eq:case-b}\end{equation}
where $z_{\pm}=b_{1}\Delta k_{x}+b_{3}\Delta 
k_{y}\pm ib_{2}\Delta k_{y}$ with real constants
$b_1$, $b_2$ and 
$b_3$. 
Note that in materials with e.g. 3- or 4-fold
rotational symmetry within $xy$ plane,
one has $b_{1}=b_{2}$ and $b_{3}=0$, leading to $z_{\pm}\propto 
\Delta  k_{x}\pm i\Delta k_{y}$.
Thus we have shown that 
the generic Hamiltonian with spin-orbit coupling 
with time-reversal- and inversion-symmetries 
decouples into a pair of 
Hamiltonians describing two-component fermions with opposite sign of the 
corresponding mass terms. 
The eigenenergies are
$E=E_{0}\pm \sqrt{m^{2}+z_{+}z_{-}}$. 
The inversion matrix in this basis 
is written as 
$\eta_a\otimes \tau_z=\eta_a\mathrm{diag}(1,-1,1,-1)$. 

In 3D, by the similar analysis, we can write down 
the generic form of the effective Hamiltonian. 
If we assume an additional symmetry of e.g. 3- or 4-fold
rotational symmetry within $xy$ plane, the Hamiltonian 
simplifies to 
\begin{equation}
{H}({\bf k})=E_{0}+\left(
\begin{array}{cccc}
m&z_{-}&& C\Delta k_z\\
z_{+}&-m&C\Delta k_z&\\
&C\Delta k_z&m&-z_{+}\\
C\Delta k_z&&-z_{-}&-m
\end{array}\right).\label{eq:3Deff}
\end{equation}
where $z_{\pm}=A(\Delta k_{x}\pm i \Delta k_{y})$ with real constants
$A$ and 
$C$, after unitary transformations. By a unitary transformation, 
this is the same as the Hamiltonian 
for Bi$_2$Se$_3$ discussed in \cite{Zhang09} up to the linear order in 
$\Delta k$. This model is used in \cite{Shindou08,Shindou10} to 
discuss disorder effects on the 3D topological insulators.

\subsection{Inversion asymmetric systems}
\label{sec:I-asym}In the inversion-asymmetric systems,
the phase transition is quite different from the inversion-symmetric systems.  
First we show that the gap cannot close at the TRIMs ${\bf k}={\bf \Gamma}_{i}$. 
This is because at the TRIMs, all the states are doubly degenerate, and the codimension 
is five \cite{Avron88,Avron89}.
To see this explicitly, we consider $4\times 4$ 
Hamiltonian matrix with the constraint (\ref{time-reversal}),
as the number of states involved is four, 
It leads to a result
\begin{equation}
{H}({\bf k}={\bf \Gamma}_{i})=E_{0}+\sum_{i=1}^{5}a_{i}
\Gamma_{i}
\label{eq:asym-Gamma-i}
\end{equation}
where $a_{i}$'s and $E_{0}$ are real.
Its eigenenergies are given by 
$E_{0}\pm\sqrt{\sum_{i=1}^{5}a_{i}^{2}}$. 
The gap between the two (doubly-degenerate) bands close when 
$a_{i}=0$ for $i=1,\cdots,5$, 
but they cannot be satisfied simultaneously 
by tuning only one parameter $m$. (We note that
${\bf k}$ is fixed to be ${\bf k}={\bf \Gamma}_{i}$.)
Therefore, the gap does not close at the TRIMs ${\bf k}={\bf \Gamma}_{i}$.

If ${\bf k}$ is not among the TRIMs,
each band is {\it non-degenerate}. We thus retain only one valence 
and one conduction bands  (Therefore the states are involved and $H({\bf k}$) is set to be 2$\times$2). 
Equation (\ref{time-reversal}) gives 
no constraint for the Hamiltonian near ${\bf k}$. 
The Hamiltonian near ${\bf k}$ thus reduces to 
\begin{equation}
H=\left(
\begin{array}{cc}
a & c\\
c^{*} & b
\end{array}
\right),
\end{equation}
where
$a$, $b$ are real functions of ${\bf k}$ and $m$, and $c$ is 
a complex function of ${\bf k}$ and $m$. 
The gap closes when the three conditions 
\begin{equation}
a=b,\ {\rm Re}c=0,\ {\rm Im}c=0,
\label{eq:gapclosingcond}
\end{equation}
are satisfied,
i.e. the codimension is 3 \cite{vonNeumann29,Herring37}.

\begin{figure}
\includegraphics[scale=0.119]{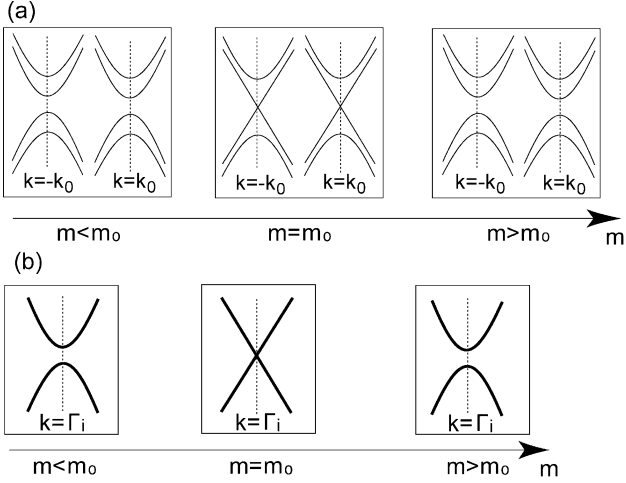}
\caption{Generic gap-closing in 2D for (a) inversion-asymmetric and 
(b) inversion-symmetric cases. In case (b) all the states are
doubly degenerate by Kramers theorem.}
\label{fig:degeneracy}\end{figure}

In 2D there are three variables $k_x$, $k_y$ and $m$, which is
the same number as the gap-closing conditions; therefore, the gap can close at some ${\bf k}$($\neq {\bf \Gamma}_i$) 
when the parameter $m$ is tuned. Let $m_0$ to be the value of $m$ at the 
gap closing.
Because of the time-reversal-symmetry, the gap closes simultaneously at ${\bf k}=
\pm {\bf k}_{0}$, at $m=m_{0}$, as depicted in 
Fig.~\ref{fig:degeneracy} (a).
In 2D, the effective Hamiltonian 
near the gap-closing point ${\bf k}={\bf k}_{0}(\neq {\bf \Gamma}_{i})$
reduces to 
\begin{equation}
{\cal H}=E_{0}(m,k_x,k_y)\pm (m-m_0)\sigma_z
+(k_x-k_{x0})\sigma_x+(k_y-k_{y0})\sigma_y
\label{eq:Weyl}
\end{equation}
after unitary and scale 
transformations \cite{Murakami07a}. 
Thus, to summarize our theory in the 2D systems,
generic gap-closings are
classified into two cases shown in Fig.~\ref{fig:degeneracy} (a)(b) for inversion-asymmetric and inversion-symmetric cases 
respectively. 
Other types of gap-closings are prohibited because of level repulsion between 
the states in the 
valence and the conduction bands by tuning a single parameter $m$.

In 3D, in contrast with the 2D case, there are 4 variables
 $m,k_x,k_y,k_z$, and the situation is different.
Because the gap-closing condition consists of three equations,
it determines a curve 
in the four-dimensional  $(k_x,k_y,k_z,m)$-hyperspace, 
which we call a ``string'' $C$.
Generally, this string $C$ in $(k_x,k_y,k_z,m)$-hyperspace occupies a finite region in 
$m$-direction $m_1\leq m\leq m_2$. 
Within this region $m_1\leq m\leq m_2$ the bulk gap is closed and the 
system is in a 
gapless phase.
As we see in the subsequent discussion, this gapless phase is
a topological phase, protected by the 
existence of monopoles and antimonopoles in ${\bf k}$ space.

\subsubsection{Monopole-antimonopole pair creation and annihilation 
in ${\bf k}$ space}
We note that we are focusing on the 3D inversion-asymmetric systems. To see the behavior of the gap-closing points in ${\bf k}$ space, 
we note that each gap-closing point carries a topological number. 
Such gap-closing point is regarded as 
a monopole or an antimonopole in ${\bf k}$ space \cite{Berry84,Volovik87,Volovik01,Volovik,Murakami03b}. 
It can be seen by associating 
the Bloch wavefunctions with nondegenerate spectrum 
to U(1) gauge field in the ${\bf k}$ space;
\begin{eqnarray}
{\bf A}_{n}({\bf k})&=&-i\langle u_{{\bf k}n}|\nabla_{{\bf k}}|u_{{\bf k}n}\rangle,\\
{\bf B}_{n}({\bf k})&=&\nabla_{{\bf k}}\times {\bf A}_{n}({\bf k}),\\
\rho_{n}({\bf k})&=&\frac{1}{2\pi}\nabla_{{\bf k}}\cdot {\bf B}_{n}({\bf k}).
\end{eqnarray}
where $|u_{{\bf k},n}\rangle$ is the Bloch wavefunction for an 
$n$-th band.
 ${\bf A}_{n}({\bf k})$, ${\bf B}_{n}({\bf k})$ are called Berry connection 
and 
Berry curvature, respectively. $\rho_n({\bf k})$ is called a monopole density. 
$\rho_n({\bf k})$ vanishes when the $n$-th band is not degenerate with 
other bands, because $\nabla_{{\bf k}}\cdot (\nabla_{{\bf k}}\times \ )=0$. 
On the other hand, $\rho_n({\bf k})$ does not vanish
when the $n$-th band 
touches with another band in energy at some
${\bf k}$-point ${\bf k}={\bf k}_0$. 
In such case, the Bloch wavefunction cannot be expressed as 
a single continuous function around ${\bf k}={\bf k}_0$;
 ${\bf k}$ space should be patched with more than 
one continuous wavefunctions \cite{Kohmoto85}, 
as is similar to the vector potential around the Dirac monopole \cite{WuYang}. 
As a result, $\rho_n({\bf k})$ is shown to have a $\delta$-function singularity at at ${\bf k}={\bf k}_0$ \cite{Berry84,Volovik87,Volovik01,Volovik,Murakami03b}. 
For example, 
for the gap closing at ${\bf k}={\bf k}_{0}$ with 
linear dispersion (Weyl fermion)
\begin{equation}
{\cal H}=E_0({\bf k})+\sum_{i=1}^{3}f_i({\bf k})\sigma_i
\label{eq:Weyl2}
\end{equation}
where $f_i({\bf k}={\bf k}_{0})=0$, the monopole density 
for the lower band
is given by $q\delta({\bf k}-{\bf k}_0)$, where $q={\rm sgn}({\rm det}(\frac{\partial f_i}{\partial k_j})_{ij})|_{{\bf k}={\bf k}_{0}}$
($=\pm 1$) is called monopole charge.
In general the monopole density has the form $\rho({\bf k})=\sum_{l}q_{l}
\delta({\bf k}-{\bf k}_{l})$ where the monopole charge
$q_{l}$ is quantized to be an integer.
It is shown that when we vary the system by changing a parameter continuously, 
the monopole charge is conserved, because of the quantization of the 
monopole charge. 
The only chance for the monopole charge to change 
is a pair creation or a pair annihilation of a monopole ($q_{l}=1$) and
an antimonopole ($q_{l'}=-1$). 
\begin{figure}
\includegraphics[scale=0.142]{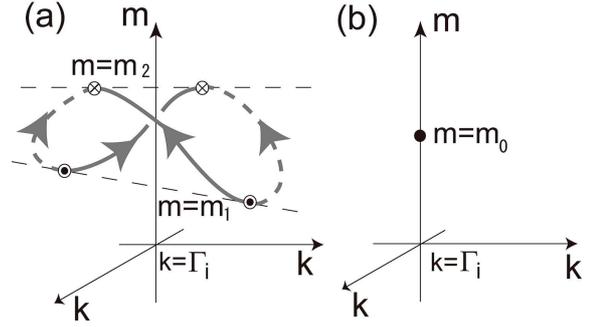}
\caption{ 
Trajectory of the gap-closing points for (a) inversion-asymmetric and 
(b) symmetric systems. For (b) inversion-symmetric systems, the gap-closing
point is located at ${\bf k}={\bf \Gamma}_i$, and is an isolated point in the
$m$-${\bf k}$ space. Only 
at $m=m_0$ the system is gapless. 
For (a) inversion-asymmetric systems, on the other 
hand, the gapless points are created in monopole-antimonopole pairs
at $m=m_1$, and
move in ${\bf k}$-space as $m$ is varied. Solid and broken curves 
represent the trajectories for monopoles and antimonopoles, respectively. 
These two curves together form a closed loop, which is the string $C$ in 
\S \ref{sec:I-asym}.
The system opens a gap only 
by pair annihilation of these gapless points at $m=m_2$. }
\label{fig:monopole}\end{figure}

\begin{figure}[h]
\includegraphics[scale=0.146]{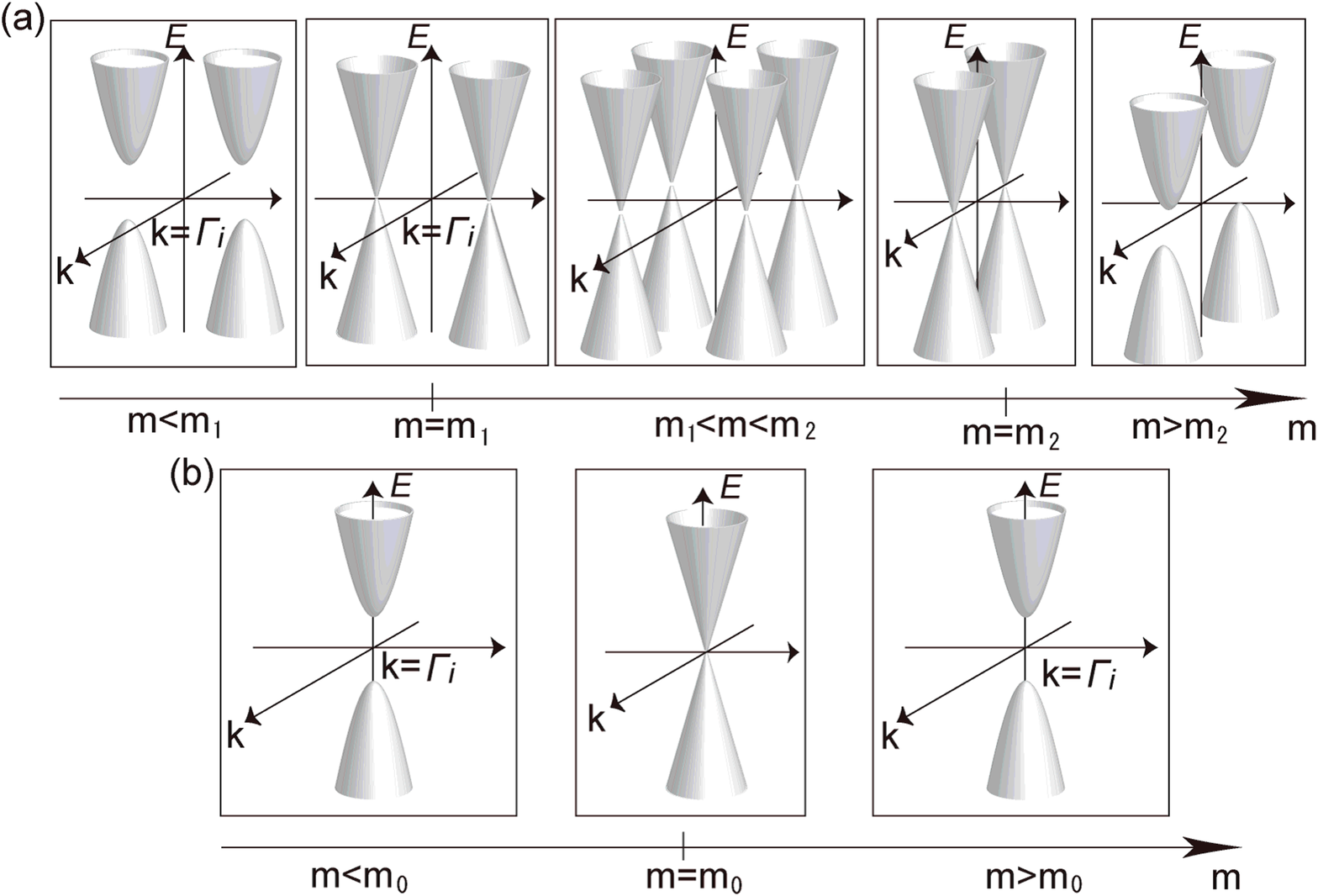}
\caption{Phase transition in 3D between 
the quantum spin Hall (QSH) and insulating phases
 for (a) inversion-asymmetric and 
(b) inversion-symmetric cases. In the case (b) all the states are
doubly degenerate.}
\label{fig:degeneracy3D}\end{figure}

In the present problem of nonmagnetic insulators, because of the time-reversal-symmetry, the
distribution of monopole charges is symmetric with respect to 
${\bf k}={\bf \Gamma}_i$: $\rho_{\alpha}({\bf k})
=\rho_{\bar{\alpha}}(-{\bf k})
=\rho_{\bar{\alpha}}(2{\bf \Gamma}_i-{\bf k})$, 
where $\bar{\alpha}$ is the label which is a time-reversed label from $\alpha$.
From these arguments we can see that the simplest case of the gap closing 
is as shown in Fig.~\ref{fig:monopole}(a) \cite{Murakami07b,Murakami08}.
Two monopole-antimonopole
pairs are created at ${\bf k}=\pm{\bf k}_{0}+{\bf \Gamma}_i$ (${\bf k}_{0}\neq 0$)
simultaneously when $m=m_1$, and the system becomes gapless. 
When $m$ is increased further, the monopoles and the antimonopoles 
move in the ${\bf k}$ space, while  
the distribution of the monopole charges remains symmetric with respect
to ${\bf \Gamma}_{i}$. 
This system can open a gap again only when all 
the monopole and antimonopole
annihilate in pairs. This occurs at $m=m_{2}$ as shown in 
Fig.~\ref{fig:monopole} (a). 
Thus the overall feature of the phase transition is
schematically expressed as in Fig.~\ref{fig:degeneracy3D}.
As the monopoles and antimonopoles are gap-closing points in ${\bf k}$ space,
 the trajectory of the monopoles and 
antimonopoles is nothing but the string $C$ (see \S \ref{sec:I-asym}.), describing the set of parameters
$(k_x,k_y,k_z,m)$ which satisfy the gap-closing conditions (\ref{eq:gapclosingcond}). 
Because the monopole charge is conserved,
the monopoles and antimonopoles are created and annihilated 
only in pairs, which means that 
the trajectory $C$ of the monopoles and 
antimonopoles form a closed loop in the $({\bf k},m)$ space. 
Namely, the string $C$ has no end point, 
because an end point of $C$ 
would violate the conservation
of monopole charge.
When $m_1<m<m_2$ the system is gapless because there 
are monopoles and antimonopoles, which are gap-closing points (see Fig.~\ref{fig:monopole}(a)).

\section{Change of the $Z_2$ topological number at the gap-closing point}
In \S 2 we classify all the generic gap-closings by tuning a single parameter
in nonmagnetic insulators. In this section we relate this to topological insulators. We will see that all these gap-closings
accompany the change of the $Z_2$ topological number and 
thus describe phase transition between the QSH phase and the ordinary insulator phase. As a result the universal phase diagram is as shown in Fig.~\ref{fig:phase-diagram-3D}, which is the main result of the paper.
\subsection{$Z_2$ topological numbers: review}
We first review the expressions for the $Z_2$ topological number given in 
\cite{Fu06a,Fu06c,Fu07a},
both in 2D and in 3D. We assume that the spectrum of the Hamiltonian has a gap,
within which the Fermi energy $E_F$ is located.
From the Kramers theorem for time-reversal-symmetric systems, 
{\it if inversion-symmetry is broken}, double degeneracy occurs only at the four TRIMs, 
and non-degenerate at other 
points.
In such systems, the $Z_2$ topological number $\nu$ 
is determined as
\begin{equation}
(-1)^{\nu}=\prod_{i=1}^{4}\delta_{i},
\ \ 
\delta_{i}=\frac{\sqrt{{\rm det}[w({\bf \Gamma}_{i})]}}{{\rm Pf}[w({\bf \Gamma}_{i})]}
=\pm 1.
\label{eq:pfaffian}
\end{equation}
Here $w({\bf k})$ is a unitary matrix with elements given by
$w_{mn}({\bf k})=\langle u_{{\bf -k},m}|\Theta|u_{{\bf k},n}\rangle$, and 
$|u_{{\bf k},n}\rangle$ is the Bloch wavefunction of an 
$n$-th band whose eigenenergy lies below
$E_F$.
The branch of the square root of the determinant is so chosen 
that the wavefunctions (including their phases) 
are continuous in the whole Brillouin zone.
On the other hand, {\it in inversion-symmetric systems}, the formula simplifies 
drastically; it is given by
\begin{equation}
(-1)^{\nu}=\prod_{i=1}^{4}\delta_{i}, \ \ 
\delta_{i}=\prod_{m=1}^{N}\xi_{2m}({\bf \Gamma}_{i}),
\label{eq:parity}
\end{equation}
where 
$\xi_{2m}({\bf \Gamma}_i)$ ($=\pm 1$) is the parity eigenvalue of the 
Kramers pairs at the TRIM ${\bf \Gamma}_{i}$, and 
$N$ is the number of Kramers pairs below $E_F$.

In 3D, there are four 
$Z_2$ topological numbers written as 
$\nu_{0};(\nu_{1}\nu_{2}\nu_{3})$ \cite{Fu07a,Moore07}, given by
\begin{equation}
(-1)^{\nu_{0}}=\prod_{i=1}^{8}\delta_{i}, \ \
(-1)^{\nu_{k}}=\prod_{n_k=1; n_{j\neq k}=0,1}\delta_{i=(n_1n_2n_3)}. 
\label{eq:Z2-3D}\end{equation}
When $\nu_0=0$ it is called the weak topological insulator (WTI), 
while when  $\nu_0=1$ it is called the strong topological insulator (STI). 
These topological numbers in 3D determine the topology of the 
surface states for arbitrary crystal directions \cite{Fu07a}.
We note that among the four $Z_2$ topological numbers in 3D, 
only $\nu_{0}$ is robust against nonmagnetic disorder, 
while $\nu_{1}$, $\nu_{2}$, and $\nu_{3}$ are meaningful only 
for a relatively clean sample \cite{Fu07a}.

\subsection{Change of the Z$_2$ topological number at gap closing}
We have shown in \cite{Murakami07a,Murakami07b,Murakami08} that all the 
types of gap closing found in \S 2 accompany the change of the Z$_2$ topological
number, and thus entail phase transition between the QSH phase and the ordinary insulator
phase. Following \cite{Murakami07a,Murakami07b,Murakami08} we explain 
its proof and its implications. 

For the inversion-symmetric systems in 2D and 3D, from Eq.~(\ref{eq:Z2-3D}), 
the Z$_2$ topological numbers are 
given by the parity eigenvalues of the occupied states. 
The gap at ${\bf k}={\bf \Gamma}_i$ collapses  when $m=0$. Hence only 
 the parity eigenvalue at ${\bf k}={\bf \Gamma}_i$ can change at the phase transition. 
Since the inversion matrix is given by ${P}=
\eta_{a}\otimes\tau_z=\eta_{a}\sigma_{0}\otimes\tau_z$, 
the parity eigenvalues of (\ref{eq:case-b}) at ${\bf k}={\bf \Gamma}_i$ are $-\eta_{a}(=\eta_{b})$ and $+\eta_{a}$ for 
the lower-band 
states at $m>0$ and $m<0$, respectively. Hence, the parity 
eigenvalue changes sign, and the $Z_2$ topological number $\nu$ 
changes by one. 
Thus, on the two sides of the gap closing, $m>0$ and $m<0$, 
one of the phases is the QSH phase, while the other one is the ordinary 
insulator phase.

For the inversion-asymmetric 2D systems, the homotopy characterization in 
Ref.~\cite{Moore07} 
is applicable; for the lower band at the critical value $m=m_0$, 
there is one vortex at ${\bf k}_{0}$ and one antivortex at 
$-{\bf k}_{0}$. Thus, when the parameter $m$ is tuned across $m=m_0$, 
the Chern number for the whole contracted surface \cite{Moore07}
changes by one. Thus, the $Z_2$ topological 
numbers are different by one for the $m>m_0$ and  the $m<m_0$ sides.
One of the phases is the QSH phase, while the other is the ordinary insulator.
\begin{figure}
\includegraphics[scale=0.117]{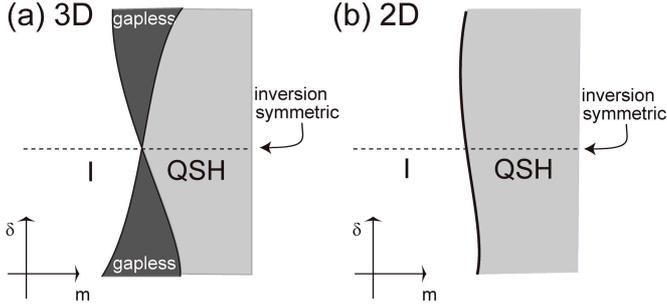}
\caption{Universal phase diagram for the quantum spin Hall (QSH) and insulator (I)
phases in (a) 3D and (b) 2D. 
$m$ is a parameter driving the phase transition, and $\delta$ represents
inversion symmetry breaking. $\delta=0$ corresponds
to the inversion-symmetric system.} 
\label{fig:phase-diagram-3D}\end{figure}

For the inversion-asymmetric 3D case, 
we can relate the shape of the trajectory (``loop'') 
of the gap closing points  (Fig.~\ref{fig:monopole}(a)) with
the change in the topological number. It has been 
discussed in \cite{Murakami08}. The rule is simple;
if the loop winds around a TRIM ${\bf \Gamma}_i$ once (as in Fig.~\ref{fig:monopole}(a)), 
the index $\delta_i$ 
changes between the phase in $m>m_2$ and that in $m<m_1$. The change of the
$Z_2$ topological numbers $\nu_0;(\nu_1\nu_2\nu_3)$ follows from it. 
In particular, $\nu_0$ changes when the loop winds around only one of
the TRIMs as in Fig.~\ref{fig:monopole}(a). 
It is reasonable since by restoring the inversion symmetry via 
continuous change of the Hamiltonian,  
Fig.~\ref{fig:monopole}(a) should reduce to Fig.~\ref{fig:monopole}(b).

To summarize the universal phase diagram is obtained as
Fig.~\ref{fig:phase-diagram-3D}. We note that for inversion-asymmetric 
3D systems there should be a finite region of ``topological" gapless phase
between the topological and trivial phases. This is exactly the region where 
the string $C$ lies (see Fig.~\ref{fig:monopole}(a)).

\section{Examples of the gap closing: models and materials}
\subsection{2D systems}
The Kane-Mele model \cite{Kane05a} on the honeycomb lattice is studied within 
our theory and it agrees with our results. 
The hopping $t$ term and the spin-orbit $\lambda_{\rm SO}$ term 
preserves the inversion symmetry. Nevertheless, 
the phase diagram
(Fig.~1 in \cite{Kane05a}) is in the $\lambda_v$-$\lambda_R$ space,
and both $\lambda_v$ and $\lambda_R$ terms break inversion symmetry. 
Thus this model in $\lambda_v$-$\lambda_R$ space falls into the class of inversion-asymmetric 2D model 
(Fig.~\ref{fig:degeneracy} (a)). 
As we have predicted, the gap then should close at non-TRIM points.
In accordance with this expectation, 
the gap closes at the $K$ and $K'$  points (i.e. non-TRIM) at the phase transition.

In the CdTe/HgTe/CdTe quantum well, the 2D quantum 
spin Hall state has been observed experimentally \cite{Bernevig06f,Konig07}. 
By increasing the well thickness $d$ over the critical thickness
$d_c\sim64$\AA\  the system undergoes a transition from the ordinary insulator
to the QSH system. The gap closing here is the 
class of 2D inversion symmetric system in Fig.~\ref{fig:degeneracy}(b), 
and should occur at one of the TRIMs. 
Indeed, when $d=d_c$, the bulk gap closes at the $\Gamma$ point (i.e. TRIM), and 
the parity eigenvalues are exchanged. 
Moreover, 
the effective model is discussed in \cite{Bernevig06f} up to the $k^2$ order.
We can compare it to our effective theory (\ref{eq:case-b}) and they are 
in complete agreement to the linear order in $k$. 

\subsection{3D systems}
\begin{figure}
\includegraphics[scale=0.18]{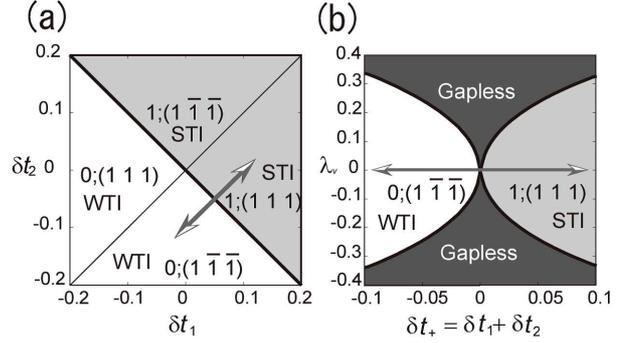}
\caption{ 
Phase diagrams for the Fu-Kane-Mele model with $\delta t_3=0$,
$\delta t_4=0$. $t_1$ and $t_2$ are the bonds along the [111] and 
[$1\bar{1}\bar{1}$] directions. 
We put $\lambda_{\mathrm{SO}}=0.1t$. The axes are in the unit
of $t$. (a) The phase diagram in 
$\delta t_1$-$\delta t_2$ plane obtained in Ref.~\cite{Fu07a}.
$\lambda_v$ is set as zero.
(b) The phase 
diagram in the $\delta t_{+}$-$\lambda_v$ plane.
 We have introduced $\lambda_v$ into the Fu-Kane-Mele model. Here 
$\delta t_{+}=\delta t_{1}+\delta t_{2}$ is changed,
while we fix $\delta t_{-}=\delta t_{1}-\delta t_{2}=0.1t$. 
The arrows in (a) and (b) correspond to the identical
change in parameters. }
\label{fig:phase-dia1}
\end{figure}

We take the 3D tight-binding model proposed by 
Fu, Kane and Mele \cite{Fu07a} on a diamond lattice as an 
example. This model
shows a transition between STI and WTI.
This model 
is 3D inversion symmetric, and the gap should close at one of the TRIM. Indeed, as calculated in \cite{Fu06a},
the band gap closes at one of the X points.
Furthermore, one can use this model to confirm the universal phase diagram
in Fig.~\ref{fig:phase-diagram-3D}, including the 
topological gapless phase for inversion-asymmetric 
3D systems. We explain it, following the analysis in Ref~\cite{Murakami08}. 
 We break inversion symmetry by adding a staggered on-site 
potential $\lambda_v$, and calculate the phase diagram.
The results are shown in  Fig.~\ref{fig:phase-dia1}(b),
where the vertical axis represents the staggered on-site potential,
corresponding to the parameter $\delta$ in Fig.~\ref{fig:phase-diagram-3D}(a),
Thus we can see that if we break the inversion symmetry
there occurs a finite region of a gapless phase in the phase diagram.
These are in accordance with our theory.
Furthermore, 
The trajectory of the gap closing points (not shown in this paper, details
are in \cite{Murakami08}) agrees with Fig.~\ref{fig:monopole}(a).
Thus we have seen that the universal phase diagram in 
Fig.~\ref{fig:phase-diagram-3D} holds true in the Fu-Kane-Mele model. 

Bi$_2$Se$_3$ and Bi$_2$Te$_3$ \cite{Zhang09,Xia09} are topological insulators under intensive experimental and theoretical research. Their bulk bands have a direct gap at the $\Gamma$ point. The highest valence band and the lowest conduction band
have the opposite parities at the $\Gamma$ point, and if the spin-orbit coupling 
is weakened their energies will be reversed and they become ordinary insulators.
In other words, topological insulator phases in these materials are due to 
the gap closing at the $\Gamma$ point by making the spin-orbit coupling larger. 
These are inversion-symmetric, and are classified to the 3D inversion-symmetric 
systems, whose gap closing occurs only at the TRIMs, in agreement with the above discussion.
The effective model is constructed in \cite{Zhang09}, and is in agreement with 
our theory in (\ref{eq:3Deff}) up to linear order in ${\bf k}$.

Both in 2D and in 3D, bismuth \cite{Murakami06b,Wada10}, antimony and their alloys \cite{Hsieh08} also are interesting materials for topological
insulators \cite{Fu07a}.
The 3D Bi has $\nu=$0, 
while the 3D Sb has $\nu=$1.
This is opposite to the case of the 2D \cite{Murakami06b}. 
(Note that although Bi and Sb are semimetals, there are direct gap everywhere in the Brillouin zone. 
This makes $Z_2$ topological invariants well-defined, by artificially making the bands to 
be fully gapped by continuous deformation of the bands to eliminate the 
small carrier pockets.)
To resolve this,  
numerical analysis was performed in \cite{Fukui06b}
by artificially changing the interlayer hopping
weakened by a factor $f$ ($0<f<1$).
Between $f=1$ (3D limit) and $f=0$ (2D limit), the gap closes at several parameter values of $f$, and this is described by the 
gap closing in inversion-symmetric 3D systems. The gap closes only at the TRIMs, in accordance with our theory.



\section{Conclusions and Discussions}
\label{sec:4}
To summarize, we analyze gap closings in 2D and 3D nonmagnetic insulators by 
tuning a parameter. 
We obtain the universal phase diagrams in 2D and 3D, as schematically 
shown in Fig.~\ref{fig:phase-diagram-3D}, in a plane of the control parameter
$m$ and another parameter $\delta$ 
representing an inversion-symmetry breaking. 
The
overall behavior of gap closing is schematically shown in 
Fig.~\ref{fig:degeneracy} for 2D and  
Fig.~\ref{fig:degeneracy3D} for 3D. 

We find that in inversion-asymmetric 3D systems, there lies a finite region 
of the gapless phase between the quantum 
spin Hall and ordinary insulator phases. 
We checked this for the Fu-Kane-Mele model.
We described the phase transition in terms of the motion of the 
gap-closing points (i.e. monopoles and antimonopoles) in ${\bf k}$ 
space. The gapless phase in the inversion-asymmetric 3D system 
originates from the conservation of ``monopole charge''. 
These cases of gap-closing 
exactly coincide with the phase transitions between 
the QSH and the insulating phases. In this sense our theory characterizes the 
QSH phase from the local features in ${\bf k}$ space.
All the known models exhibiting phase transition between 
the two phases are special cases of this general classification.
To confirm this scenario, experimental pursuit of 3D inversion-asymmetric topological insulators will be interesting and important.

\section*{Acknowledgment}
The author would like to thank 
Y.~Avishai, 
S.~Iso, S.~Kuga, N.~Nagaosa, and M. Onoda for fruitful 
discussions and helpful comments.
This research is supported in part 
by Grant-in-Aid for Global COE Program 
"Nanoscience and Quantum Physics"
from the Ministry of Education,
Culture, Sports, Science and Technology of Japan.






\begin{thebibliography}{99}
\bibitem{Murakami03a}
S. Murakami, N. Nagaosa, and S.-C. Zhang, Science \textbf{301}, 1348 (2003).
\bibitem{Sinova04}
J. Sinova, D. Culcer, Q. Niu, N. A. Sinitsyn, T. Jungwirth, and A. H. MacDonald, Phys. Rev. Lett. \textbf{92}, 126603 (2004).
\bibitem{Kane05a}
C. L. Kane and E. J. Mele, Phys. Rev. Lett.
\textbf{95}, 146802 (2005).
\bibitem{Kane05b}
C. L. Kane and E. J. Mele, Phys. Rev. Lett.
\textbf{95}, 226801 (2005).
\bibitem{Bernevig05a}
B. A. Bernevig and S.-C. Zhang,
Phys. Rev. Lett. \textbf{96}, 106802 (2006).
\bibitem{Wu05}
C. Wu, B. A. Bernevig, and S.-C. Zhang, Phys. Rev. Lett. \textbf{96}, 106401
(2006).
\bibitem{Xu05}
C. Xu and J. E. Moore,
Phys. Rev. B\textbf{73}, 045322 (2006).
\bibitem{Konig07}
M. K{\"o}nig, S. Wiedmann, C. Br{\"u}ne, A. Roth, H. Buhmann, 
L. W. Molenkamp, X.-L. Qi, and S.-C. Zhang
Science \textbf{318} 766 (2007).
\bibitem{Takahashi10}
Ryuji Takahashi and Shuichi Murakami
Phys. Rev. B {\bf 81}, 161302 (R)(2010).
\bibitem{Murakami07a}
S. Murakami, S. Iso, Y. Avishai, M. Onoda, and N. Nagaosa,
Phys. Rev. B\textbf{76}, 205304 (2007).
\bibitem{Murakami07b}
S. Murakami, New J. Phys.\textbf{9}, 356 (2007); (Corrigendum) 
{\it ibid.} \textbf{10},
029802 (2008).
\bibitem{Murakami08}
S. Murakami, S. Kuga, Phys. Rev. B \textbf{78}, 165313 (2008).
\bibitem{Fu06a}
L. Fu and C. L. Kane, Phys. Rev. B\textbf{74}, 195312 (2006).
\bibitem{Fu06c}
L. Fu and C. L. Kane, 
Phys. Rev. B\textbf{76}, 045302 (2007).
\bibitem{Fu07a}
L. Fu, C. L. Kane, and E. J. Mele, Phys. Rev. Lett. \textbf{98}, 106803 (2007).
\bibitem{Moore07}
J. E. Moore and L. Balents, Phys. Rev. B \textbf{75}, 121306(R) (2007) .
\bibitem{Kobayashi07}
A. Kobayashi, S. Katayama, Y. Suzumura, H. Fukuyama, 
J. Phys. Soc. Jpn. {\bf 76}, 034711 (2007).
\bibitem{Zhang09}
H. J. Zhang, C. X. Liu, X. L. Qi, X. Dai, Z. Fang, and S. C. Zhang, 
Nature Phys. \textbf{5},438 (2009).
\bibitem{Shindou08}
R. Shindou and S. Murakami, Phys. Rev. B{\bf 79}, 045321 (2009).
\bibitem{Shindou10}
Ryuichi Shindou, Ryota Nakai, Shuichi Murakami, 
New J. Phys. 12, 065008 (2010).
\bibitem{Avron88}
J. E. Avron, L. Sadun, J. Segert, and B. Simon, 
Phys. Rev. Lett. \textbf{61}, 1329 (1988).
\bibitem{Avron89}
J. E. Avron, L. Sadun, J. Segert, and B. Simon, 
Commun. Math. Phys. \textbf{124}, 595 (1989).
\bibitem{vonNeumann29}
V. J. von Neumann and E. Wigner, Physik. Zeitschr. \textbf{30}, 467 (1929).
\bibitem{Herring37}
C. Herring, Phys. Rev. \textbf{52}, 361; \textit{ibid.} \textbf{52}, 365 
(1937).
\bibitem{Berry84}
M. V. Berry, Proc. Roy. Soc. London Ser A \textbf{392}, 45 (1984).
\bibitem{Volovik87}
G. E. Volovik, JETP Lett. {\bf 46} 98 (1987).
\bibitem{Volovik01}
G. E. Volovik, Phys. Rep. {\bf 351} 195 (2001).
\bibitem{Volovik}
G. E. Volovik, {\it The Universe in a Helium Droplet}, 
(Oxford University Press, Oxford, 2003).
\bibitem{Murakami03b}
S.~Murakami and N.~Nagaosa, 
Phys. Rev. Lett. \textbf{90}, 057002 (2003).
\bibitem{Kohmoto85}
M.~Kohmoto, Ann. Phys. \textbf{160}, 343 (1985).
\bibitem{WuYang}
T. T. Wu and C. N. Yang, Phys. Rev. D\textbf{12}, 3845 (1975).

\bibitem{Bernevig06f}
B. A. Bernevig, T. L. Hughes, S.-C. Zhang,
Science \textbf{314}, 1757 (2006).


\bibitem{Xia09}
Y. Xia, D. Qian, D. Hsieh, L. Wray, A. Pal, H. Lin, A. Bansil, D. Grauer, Y. S.
Hor, R. J. Cava, and M. Z. Hasan. Nature Phys. \textbf{5},398 (2009).
\bibitem{Murakami06b}
S. Murakami, Phys. Rev. Lett. \textbf{97}, 236805 (2006).
\bibitem{Wada10}
M. Wada, S. Murakami, F. Freimuth, G. Bihlmayer, arXiv:1005.3912.
\bibitem{Hsieh08}
D. Hsieh, D. Qian, L. Wray, Y. Xia, Y. S. Hor, R. J. Cava and 
M. Z. Hasan, Nature \textbf{452}, 970 (2008).
\bibitem{Fukui06b}
T. Fukui and Y. Hatsugai, 
J. Phys. Soc. Jpn. \textbf{76}, 053702 (2007).

\end{thebibliography}








\end{document}